\documentclass{sigchi}

\toappear{}

\usepackage{balance}  
\usepackage{graphics} 
\usepackage{txfonts}
\usepackage{times}    
\usepackage[pdftex]{hyperref}
\usepackage{color}
\usepackage{textcomp}
\usepackage{booktabs}
\usepackage{ccicons}
\usepackage{todonotes}
\usepackage{hyperref}

\makeatletter
\def\url@leostyle{%
  \@ifundefined{selectfont}{\def\UrlFont{\sf}}{\def\UrlFont{\small\bf\ttfamily}}}
\makeatother
\def\pprw{8.5in}
\def\pprh{11in}

\usepackage{varwidth}
\usepackage{xcolor}

\definecolor{linkColor}{RGB}{6,125,233}
\hypersetup{%
  pdftitle={SIGCHI Conference Proceedings Format},
  pdfauthor={LaTeX},
  pdfkeywords={SIGCHI, proceedings, archival format},
  bookmarksnumbered,
  pdfstartview={FitH},
  colorlinks,
  citecolor=black,
  filecolor=black,
  linkcolor=black,
  urlcolor=linkColor,
  breaklinks=true,
}

\makeatletter
\def\@copyrightspace{\relax}
\makeatother
\begin{document}

\title{CrowdTone: Crowd-powered tone feedback and improvement system for emails}

\numberofauthors{2}
\author{%
  \alignauthor{Rajan Vaish\\
    \affaddr{Stanford University}\\
    \affaddr{Stanford, CA, USA}\\
    \email{rvaish@cs.stanford.edu}}\\
  \alignauthor{Andr\'es Monroy-Hern\'andez\\
    \affaddr{Microsoft Research}\\
    \affaddr{Redmond, WA, USA}\\
    \email{amh@microsoft.com}}\\
}

\maketitle

\begin{abstract}
In this paper, we present CrowdTone, a system designed to help people set the appropriate tone in their email communication. CrowdTone utilizes the context and content of an email message to identify and set the appropriate tone through a consensus-building process executed by crowd workers. We evaluated CrowdTone with 22 participants, who provided a total of 29 emails that they had received in the past, and ran them through CrowdTone. Participants and professional writers assessed the quality of improvements finding a substantial increase in the percentage of emails deemed ``appropriate'' or ``very appropriate'' --- from 25\% to more than 90\% by recipients, and from 45\% to 90\% by professional writers. Additionally, the recipients' feedback indicated that more than 90\% of the CrowdTone processed emails showed improvement.
\end{abstract}

\keywords{Crowdsourcing, Amazon Mechanical Turk, Tone improvement, Email writing}

\section{Introduction}

Setting the right tone --- attitude expressed through words --- in written communications often determines whether a reader will interpret the message as the writer intended~\cite{greer2016introduction}. 

Despite changes in communication practices, email is still one of the most popular forms of communication in professional settings. Every day, over 200 billion emails are written worldwide~\cite{emailnumber}. 
Despite its popularity, email is prone to misunderstandings just like other types of written communication 
~\cite{ego, fastcompany, egomore, kruger2005egocentrism, epley2005you, byron2008carrying}. 
For example, researchers have found that email senders often have the erroneous belief that their recipients will identify the intended emotion in  their messages ~\cite{fastcompany, kruger2005egocentrism, epley2005you},  while the reality is that close to half (44\%) of email recipients fail to identify the intended tone of an email message. Furthermore, the same researchers found that when the tone of an email is unclear, recipients tend to interpret the email based on their stereotypes and existing assumptions of the writer.

There are a few tools that automatically detect tone~\cite{ibm, tonechecker} and sentiment~\cite{google, alchemy} in  writing, but in order to actually set the appropriate tone, people  hire professional copywriters and proofreaders for their important communications.


Only few people and organizations can afford their own expert writers, so platforms like Upwork.com and Wordy.com give people access to pools of professional writers on-demand. However,  these services can be costly and time-consuming (e.g. finding the right people, setting a fair contract, spending time in  back and forth communication, etc.)

Previous systems research has looked into more efficient writing solutions that rely on crowds. Work in this space has proven that, with proper scaffolding, crowds can do expert-quality work such as shortening text, correcting grammar, and finding and formatting citations~\cite{Bernstein:2010:SWP:1866029.1866078, Kim:2014:EEC:2531602.2531638, Kittur:2011:CCC:2047196.2047202, Luther:2015:SAE:2675133.2675283}. Despite all this work, no systems research has focused  on tone, and more specifically emails' tone. 

In order to address the need for tone improvement in email communication, we created CrowdTone, a crowd-powered tone-improvement system. CrowdTone first receives an email's main content, along with basic information about the sender, the receiver, and open-ended context elicited by the user. Then, CrowdTone outputs an improved version of the email, with a tone appropriate to the context and receiver. CrowdTone uses crowd workers from Amazon Mechanical Turk to take emails through a step-by-step tone-scaffolding process that identifies the original tone, improves it through consensus workflow, and outputs the best result. 

We evaluated CrowdTone with 22 participants recruited from our own organization. We asked participants to provide emails they had received that they had perceived as problematic with regards to tone. For example, some participants shared emails they received from students being rude or inappropriate. Participants were asked to remove the name of the sender and other identifiable information to maintain their anonymity. 
Based on the participants' feedback, more than 90\% of the emails were improved in tone, while the percentage of email messages deemed ``appropriate'' or ``very appropriate'' rose from about 25\% to more than 90\%. When surveyed, 75\% of these participants ``agreed'' or ``totally agreed'' when asked whether the CrowdTone emails were of high quality --- matching their writing expectations. Besides, the tone-scaffolding process was reported as easy and effective by the crowd-workers. 

The core contribution of this work is the design, implementation, and evaluation of a crowd-powered process that self-identifies --- without user instructions ---  and improves the email tone with basic context and information provided. Though outside the scope of this paper, we believe that our approach and findings can be applied to different domains and media or any form of written communications.

\section{RELATED WORK}
The need to communicate a message over email in the intended way is non-trivial  ~\cite{fastcompany, kruger2005egocentrism, epley2005you, byron2008carrying}. In fact, several online services, such as Wordzen~\cite{wordzen} and Crystal~\cite{crystal},  are trying to address this same need. Wordzen focuses on grammar improvements, while Crystal makes useful suggestions based on the recipient's personality, however none of these systems focuses on \textit{improving} the tone per se.

The importance of tone~\cite{greer2016introduction}, along with the need to understand sentiment, has attracted the attention of companies like Google~\cite{google}, IBM~\cite{ibm, alchemy} and Microsoft~\cite{microsoft}. These companies have public APIs (Application Programming Interfaces) for analyzing sentiment. Although those APIs and other machine-based analysis tools~\cite{liwc} are extremely useful for parsing large datasets, they are not generally designed for end-users, and are not specifically designed to help \textit{improve} email tone.

Crowdsourcing is another domain being explored for writing assistance. Crowd-powered projects like Soylent~\cite{Bernstein:2010:SWP:1866029.1866078}, Legion~\cite{Lasecki:2011:RCC:2047196.2047200} and Chorus~\cite{Lasecki:2013:CCC:2501988.2502057} have encouraged researchers to harness the crowd for complex tasks --- and have even been implemented on different interfaces, including smart-watches~\cite{nebeling2016wearwrite}.  Projects like Turkomatic~\cite{Kulkarni:2012:CCW:2145204.2145354}, Ensemble~\cite{Kim:2014:EEC:2531602.2531638}, Crowdforge~\cite{Kittur:2011:CCC:2047196.2047202}, and others~\cite{teevan2016supporting} have succeeded in converting macro tasks into micro chunks that makes it easier for crowd workers to accomplish expert-level tasks such as writing articles and stories. 

To accomplish those efforts, it is important for crowd-powered systems to focus on task sequencing~\cite{cai2016chain}, scaffolding, and crowd coordination. Crowd coordination techniques such as iterative and parallel contributions ~\cite{Dow:2009:EPU:1640233.1640260, Dow:2010:PPL:1879831.1879836, Dow:2012:SCY:2145204.2145355, Dow:2011:PDS:1978942.1979359} produce valuable results that have been used to build projects like CrowdCrit ~\cite{Luther:2015:SAE:2675133.2675283} and Storia ~\cite{storia}. CrowdCrit uses scaffolding and crowd coordination to provide expert-level critiques to designers. 

In developing CrowdTone, we used existing techniques from the literature, and developed our own sequencing and scaffolding approaches to identify and improve email tone.

\section{THE CROWDTONE SYSTEM}

\begin{figure}[h]
\centering
  \includegraphics[width=0.9\columnwidth]{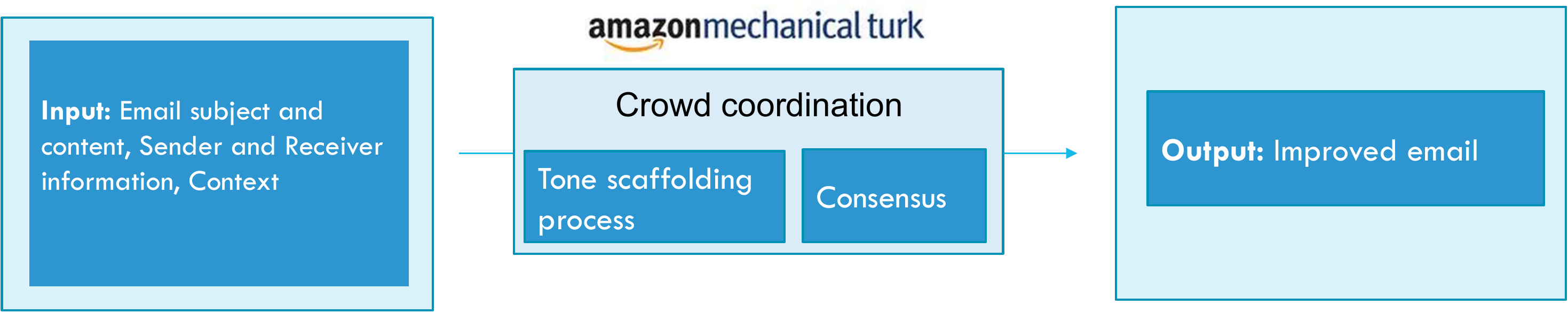}
  \caption{CrowdTone system overview: Inputs email information and context, tone scaffolding and consensus helps coordinate crowd workers from MTurk to produce improved email, improved email is produced as an output}~\label{fig:figurecrowdtone}
\end{figure}

As shown in Figure~\ref{fig:figurecrowdtone}, CrowdTone inputs email information and context, and produces an improved email via a tone-scaffolding and consensus workflow process. 

\subsection{Designing the workflow process}

The design of the process behind CrowdTone was informed by a formative study with professional writers. We recruited five North American writers from Upwork, an online freelancing platform. We selected only writers with a rating of at least 4.5 / 5 that had completed more than 100 hours of writing-related work on Upwork. We gave these writers five emails each, and asked to improve the tone and to formulate and document a step-by-step process that begins with tone identification and ends with tone improvement. 

Only two of the five writers could formulate a process. The other three found this task extremely difficult to execute. One of these three described the problem this way: 

\textit{``The reverse process is a complicated one. The rewriting of the emails was not as difficult as breaking each beat down as you requested. Kind of new territory for anyone''}

By translating our learning from this exercise and other resources~\cite{Pearl}, we explored a few pilot approaches to designing a workflow that can produce high quality output using basic context. Our approach utilized the step-by-step workflow that Upwork professionals formulated based on their experience and expertise. The rest of this section will provide a detailed description of the process --- from input through the tone scaffolding phase to output.  

\subsection{Input}
CrowdTone supports GUI and REST based input, and accepts a set of mandatory and optional information to process the email. Primarily intended for a sender of an email, CrowdTone accepts the following information: 

\begin{enumerate}
  \item Mandatory Information: provides minimal context
  \begin{enumerate}
  \item Sender: relationship (e.g. intern, student)
  \item Recipient: relationship (e.g. adviser, professor) 
  \item Email subject 
  \item Email content  
  \item Open-ended context short-description to provide information on, or the story behind the email 
  \end{enumerate}
  \item Optional Information: provides maximum context
  \begin{enumerate}
  \item Gender: sender and recipient 
  \item Native language: sender and recipient 
  \item Hierarchy relationship (e.g. professionally senior, same level, or junior), if applicable 
  \item Relationship type: friends and family, acquaintances, strangers (cold emails)
  \end{enumerate}
\end{enumerate}

If the sender decides to input mandatory information only, they are providing minimal context. However, if they decide to provide optional information, they will be providing maximum context that CrowdTone can support. 

\subsection{Crowd coordination --- phase 1: The tone scaffolding process, from tone identification to improvement}
Identifying and improving the tone of a written text is a critical and expert/professional-level task. In this section, we describe our design process and the system that produces three improved versions of  the original email, by three crowd workers. CrowdTone does not accept tone-related instruction as an input. The process focuses instead on enabling crowd workers to identify and improve the tone of an email. 

\begin{figure}[h]
\centering
  \includegraphics[width=0.9\columnwidth]{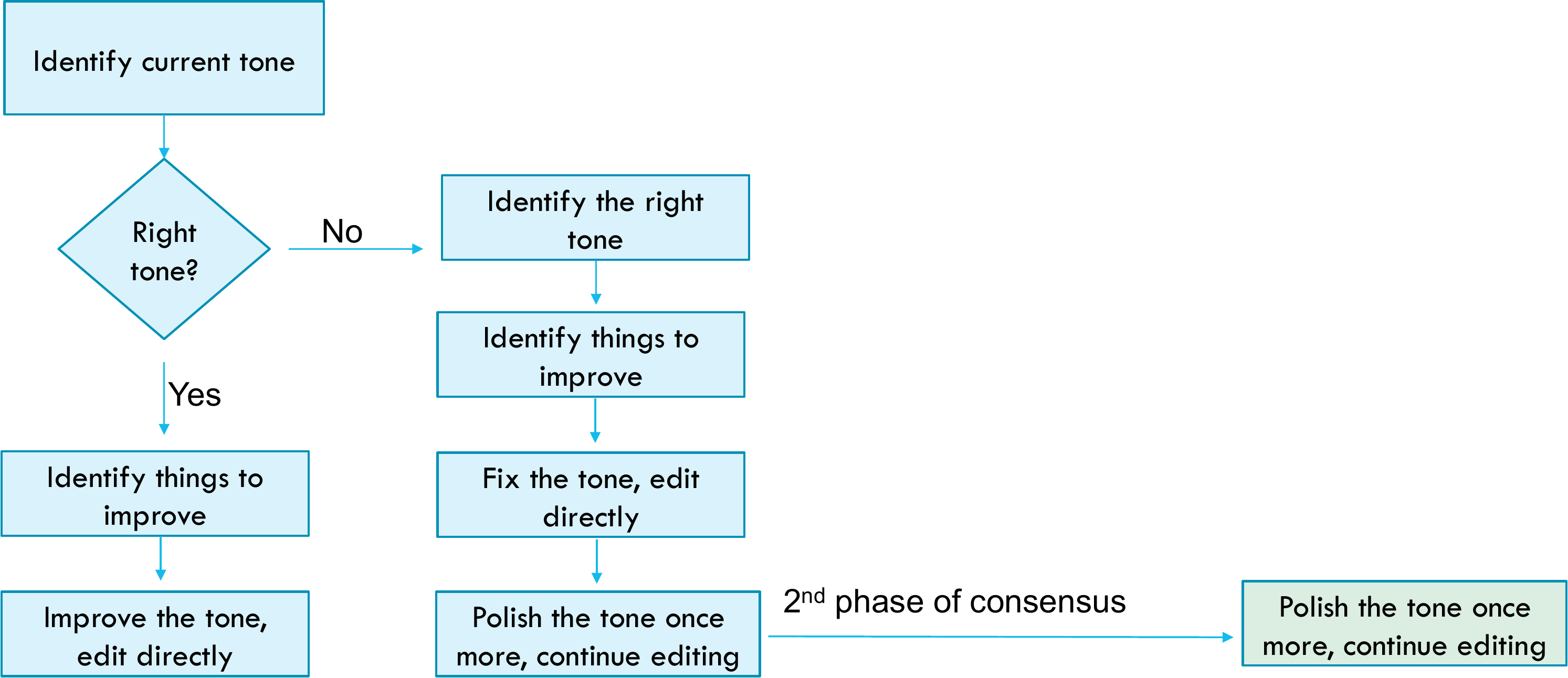}
  \caption{Tone identification to improvement process: Crowd workers go through this process to eliminate the need for tone-related instructions from requester}~\label{fig:process}
\end{figure}

As shown in the Figure~\ref{fig:process}, crowd workers begin the process by reviewing the original email and context from the perspective of its recipient. Acting from the recipient's viewpoint, the crowd workers attempt to identify the current tone using \textbf{primary} and \textbf{secondary} tone options. The two primary options are \textit{formal or informal}. The ten secondary options are --- \textit{appreciative/thankful, confident, courteous/respectful/polite, emotional/persuasive, enthusiastic/cheerful, light/humorous/friendliness, regretful/sorrowful, serious, cold/unfriendly, and enraged}. 

These options were developed after reviewing multiple articles on tone~\cite{Pearl, voiceandtone, howtotone} and running multiple pilot studies. While not exhaustive, they help the crowd workers make decisions more quickly. As one worker describes it: 

\textit{``Combining the primary and secondary tones in current and correct makes it easier to quickly bring the ideas together, enjoyed it''}

After identifying the current primary and secondary tone, each crowd worker has to decide whether that tone is right or not. Based on their yes-or-no decision, the workflow diverges in the following manner:  

\begin{figure*}[t]
  \centering
  \includegraphics[width=2\columnwidth]{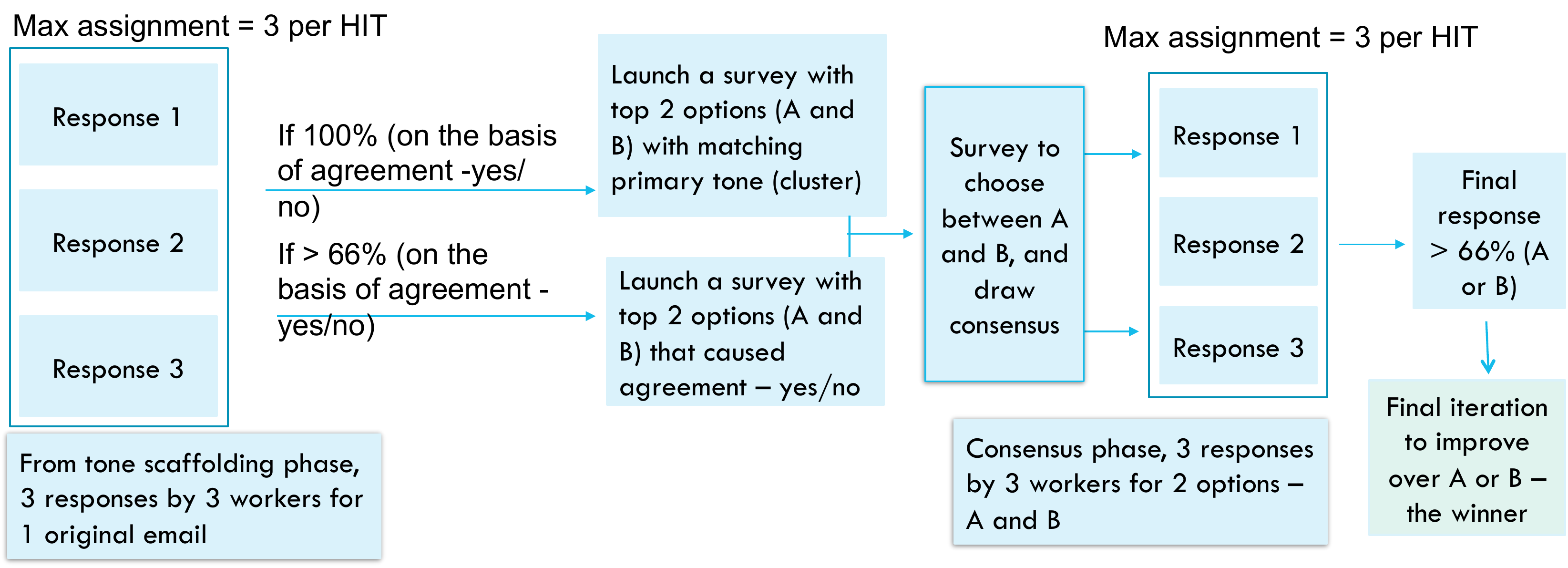}
  \caption{Workflow of the consensus approach that helps select the best improved email, and facilitates additional round of iteration to improve the best one further}~\label{fig:consensus}
\end{figure*}

\begin{enumerate}
  \item If the crowd worker chooses ``Yes'', he or she takes these actions: 
  \begin{enumerate}
  \item Step 1: Identify scope of improvement and at least one instance of text that could still be improved.
  \item Step 2: Use the suggestions from Step 1 as self-instruction and make the improvements. Here the crowd worker directly edits and improves the email. 
  \end{enumerate}

  \item If the crowd worker chooses ``No'', he or she takes these actions: 
  \begin{enumerate}
  \item Step 1: Identify the right tone by choosing the ideal primary and secondary tone options for the email and the appropriate intensity option --- \textit{very, quite close, somewhat}. For example, one can identify the tone to be \textit{very formal and appreciative}.   
  \item Step 2: Identify and list what needs to be improved to achieve the ideal tone. Here the crowd workers are encouraged to be specific and list as many instances as possible. 
  \item Step 3: Use the list and suggestions from Step 2 as self-instruction to revise and improve the email directly. 
  \item Step 4: Iterate the output from Step 3 to fine tune the email's tone. Here the crowd workers are encouraged to make further improvements through direct editing. 
  \end{enumerate}
\end{enumerate}

Overall, each email undergoes and is reviewed by three crowd-workers for confident consensus. Emails deemed initially to have correct tone goes through two crowd-worker steps, while one's with incorrect tone go through four crowd-worker steps. Eventually, producing three improved versions per original email.

This process helps improve the tone without requiring tone-related instructions from the sender.

\subsection{Crowd coordination --- phase 2: Consensus, choosing the best among the improved emails from last stage}

After getting three input-phase responses for each email, CrowdTone coordinates the crowd to get consensus on and then output the most improved email. Figure~\ref{fig:consensus} shows an overview of the consensus workflow.

\begin{figure*}[t]
  \centering
  \includegraphics[width=2\columnwidth]{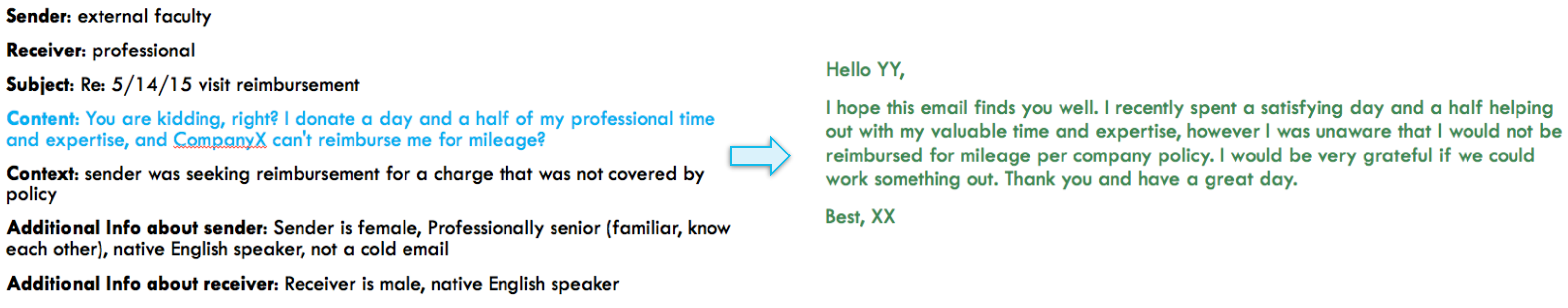}
  \caption{Before and after example: On the left we see input given to CrowdTone with additional context about sender and recipient's relationship, hierarchy and the nature of email. On the right we see the improved email that was labeled as very appropriate tone by both, recipients and professionals}~\label{fig:beforeafter}
\end{figure*}

\subsubsection{Phase A: Selecting two email versions for consensus}
After three input-phase responses are generated for each email, the next step is to select two of the three revised emails for the final consensus round. 

If ``yes'' or ``no'' were selected twice by two crowd-workers (66.66\%), the two ``yes'' or two ``no'' emails are sent forward for further consensus evaluation. 

If an email received three ``yes'' or three ``no'' responses (100\%), i.e. if an email was unanimously found to be of appropriate or inappropriate tone by three crowd workers --- the two emails to be forwarded for further consensus evaluation are chosen based on the similarity of attributes such as primary and secondary tone.

\subsubsection{Phase B: Selecting one email version for further iteration}
In the second part of the consensus phase, from the two email versions (a and b) forwarded from Phase A, three different crowd workers choose the version that --- in their view --- is the best in terms of tone, and do another iteration to improve the tone of that email. As there are two email versions, and three different crowd-workers to make the decision --- the final email selected will either enjoy a majority of 66.66\%, or 100\%. Additional iteration for improvement also brings a new perspective from another crowd member. 

\subsection{Output}

Figure~\ref{fig:beforeafter} shows a ``before'' and ``after'' example of an email that went through the CrowdTone process. By the time each email has completed the tone-scaffolding and consensus phases, six crowd workers have helped produce this output. However, depending on whether the tone of email sent was correct, the output would vary: 

\begin{enumerate}
  \item If the original email tone was deemed to be correct, the output provides: 
  \begin{enumerate}
  \item Original tone and intensity 
  \item Improved email with additional notes  
  \end{enumerate}
  \item If the original email tone was deemed to be incorrect, the output provides: 
  \begin{enumerate}
  \item Original tone and intensity 
  \item Right tone and intensity for the email
  \item Improved email with additional notes about changes and suggestions 
  \end{enumerate}
\end{enumerate}

CrowdTone supports getting input from a REST request, with each email processing generating a task id, but the requester or sender can also get a JSON format output.

\section{CROWDTONE EVALUATION}
To assess whether CrowdTone improves email tone, we conducted the four-phase evaluation described below. 

\subsection{Step 1: Understanding email usage, types and tone related use cases}

\begin{figure}[h]
\centering
  \includegraphics[width=0.9\columnwidth]{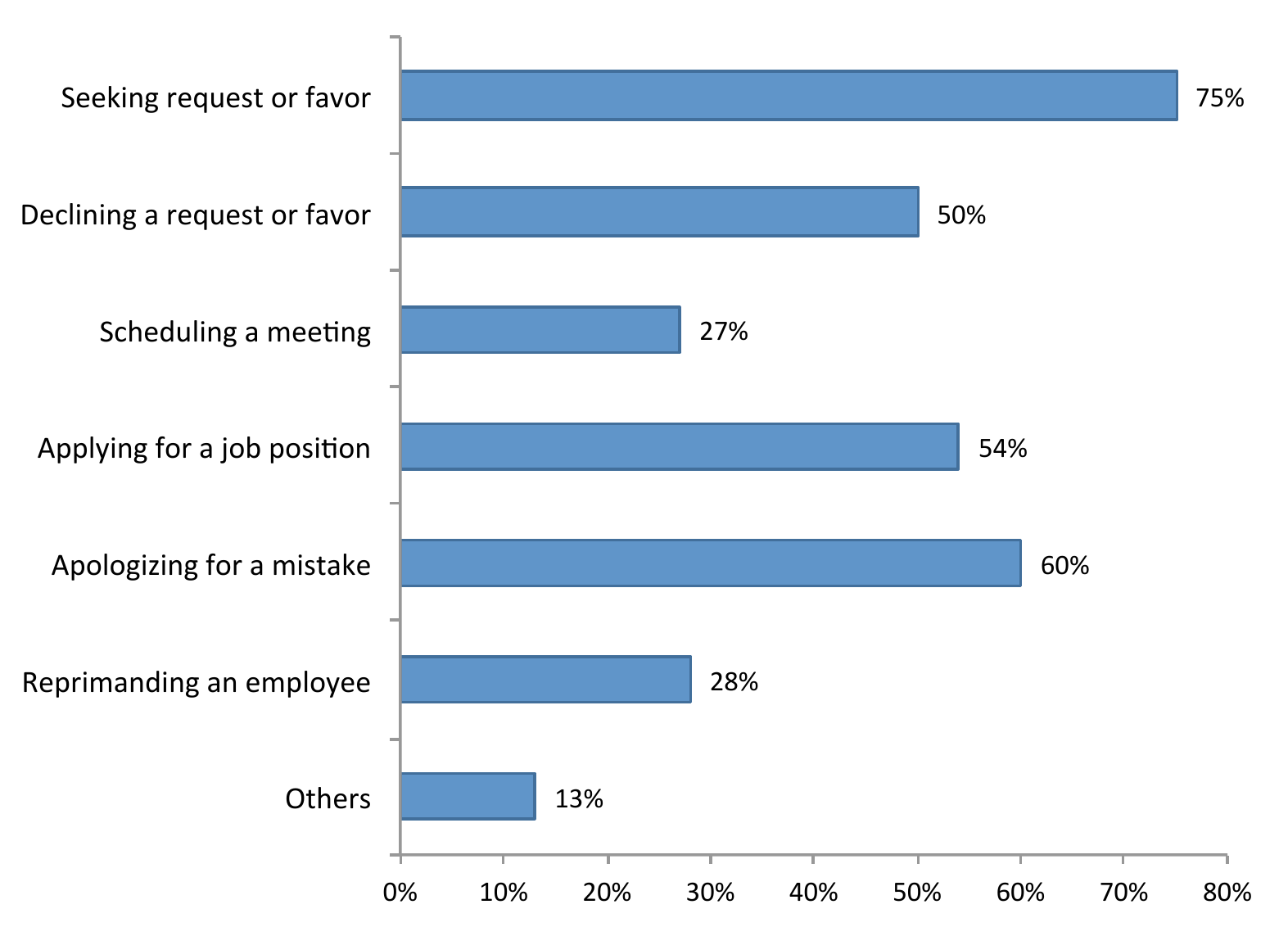}
  \caption{Email types that most need help with tone}~\label{fig:figuretypeofemails}
\end{figure}

\begin{figure}[h]
\centering
  \includegraphics[width=0.9\columnwidth]{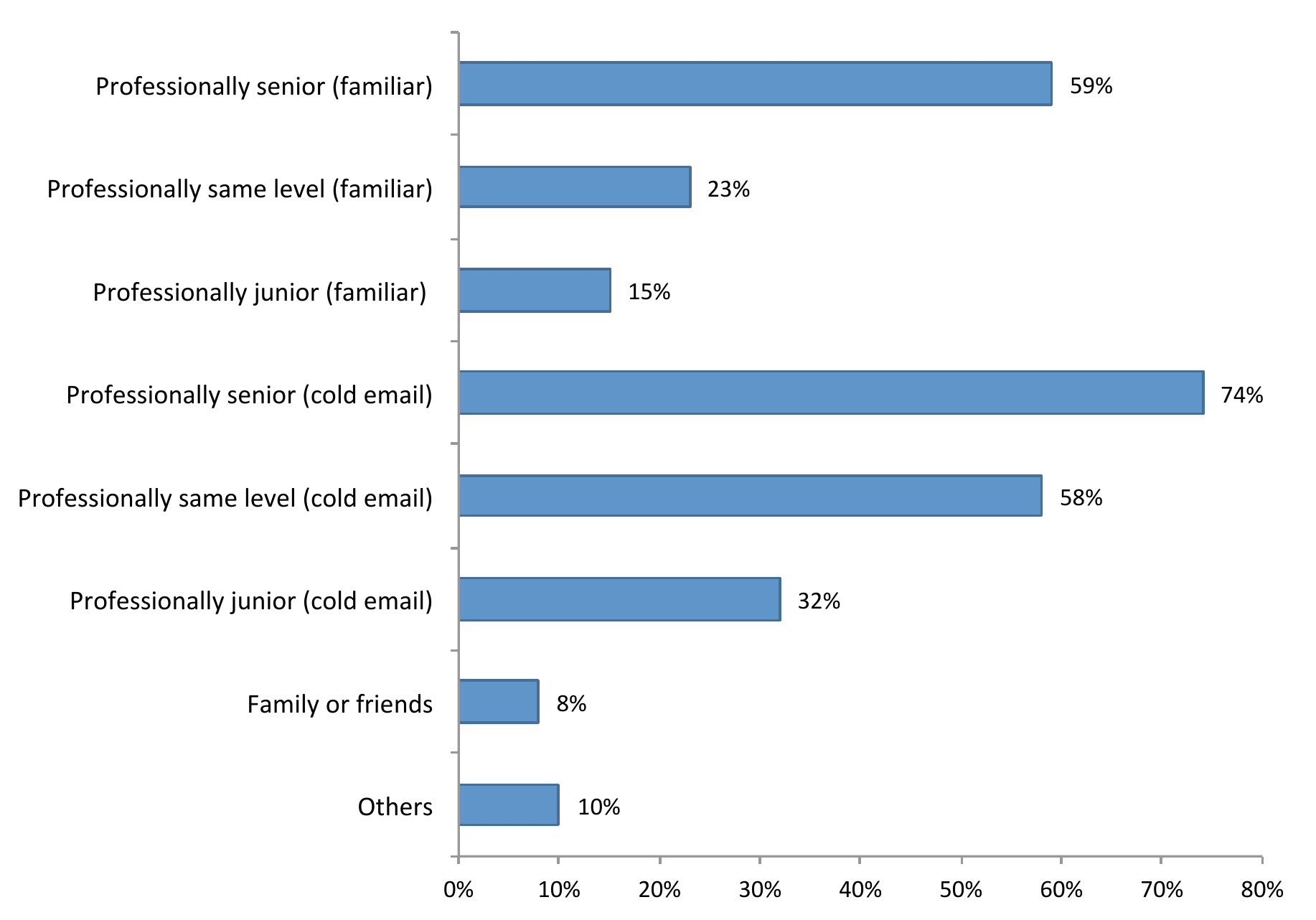}
  \caption{Category of recipients for which senders need the most help with tone}~\label{fig:figurecategoryofrecipients}
\end{figure}

To begin our evaluation, we ran a formative study within our organization to better understand people's needs when it comes to ``tone fixing'' in email. We gathered 92 responses to an online survey administered to students. Of these, 94\% of respondents thought tone to be important in email, while 60\% expressed a need for helpful feedback to improve their email tone. Of the 92 participants, 84\% assessed themselves to be fluent and expert in English and 71\% of participants were frequent email users, sending more than five emails per day. On further investigation, we found that a substantial number of respondents (74\%) felt the need for help in instances when they were  professionally emailing people more senior to them that they did not know. The expressed need for help lessened somewhat (59\%) with regard to professionally senior people they did know. Overall, most people (75\%) reported needing the most help when they were sending ``cold'' emails, that is, emails to strangers, especially when they needed to ask a favor or make a request. Figures~\ref{fig:figuretypeofemails} and \ref{fig:figurecategoryofrecipients} give an overview of these survey responses and, taken together, reflect the importance of and need for appropriate tone in emails.

\subsection{Step 2: Collect emails and initial information from the participating recipients}

After  understanding the use cases when tone-fixing is most helpful, we gathered emails received by some of our employees along with additional information on each email via a survey. We focused on the email recipients rather than senders because it is the recipients' assessment of whether a message is appropriate or inappropriate in tone that matters most~\cite{fastcompany, kruger2005egocentrism, epley2005you, byron2008carrying}. We asked employees in our organization to share an email they had received that they considered to be deficient in tone. In response, we received 29 emails sent to 22 recipients by people they knew or did not know. Before sending these emails to us, the participants first made them anonymous by removing all identifying information.

Based on a preliminary study conducted earlier with 92 employees, we learned about the type of emails where people need most help with tone fixing/improvement --- therefore, we asked our recipients to send us emails that: 

\begin{itemize}
  \item asked for a favor or made a request;
  \item constituted a job application or a query regarding job opportunities; or
  \item were apologetic or regretful. 
\end{itemize}

Among these 22 recipients, 72\% were male, while 28\% female. 35\% were native English speakers, while 86\% assessed themselves to be fluent or expert in English language. Also, 55\% participants were full time employees, while the rest were interns or temporary workers. Finally, 67\% participants were active email users and sent more than five emails per day. 

\begin{figure}[h]
\centering
  \includegraphics[width=0.9\columnwidth]{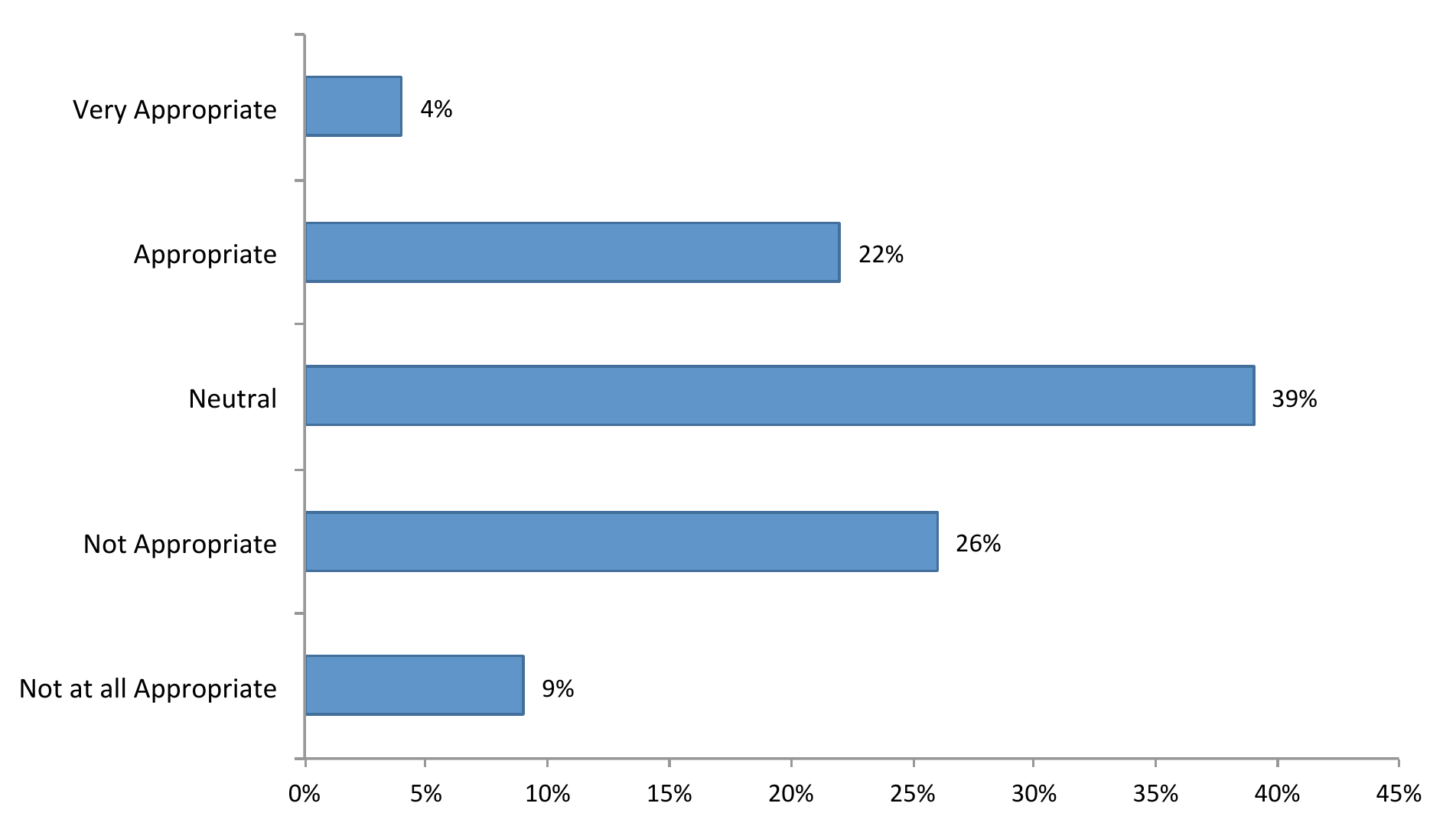}
  \caption{Appropriateness  of the tone of original emails, assessed by participating recipients}~\label{fig:approoriginal}
\end{figure}

Upon asking about the senders, we learned that 45\% were strangers to the recipients (cold emails), while they knew 52\% of them. And 41\% were professionally junior to the recipient, while 38\% were professionally same level, and 17\% were professionally senior. Based on the recipient's perception, 31\% of the senders appeared to be native English speakers and 43\% seemed to be male, while 32\% appeared female. Overall, we received a balanced set of emails.

On a scale of 1 to 5, with 1 being ``not at all appropriate'' and 5 being ``very appropriate'', we found that 7\% were reported by the recipients to have a ``not at all appropriate'' tone; 24\% to have a ``not appropriate'' tone; while 41\% were felt to be neutral. The remaining 24\% and 3\% emails were rated ``appropriate'' and ``very appropriate'' by the recipients of these emails. We can see the details in Figure~\ref{fig:approoriginal}. 

For each sample email, the participating recipients were paid \$10 after they gave feedback on the CrowdTone-improved version.

\subsection{Step 3: Run HITs (Human-Intelligent Tasks) on Mechanical Turk to get output from CrowdTone}

CrowdTone is powered by crowd workers from Amazon Mechanical Turk. In the third step of our evaluation, each email received the following two CrowdTone context treatments:

\begin{enumerate}
  \item Minimal context: mandatory open-ended context. 
  \item Maximum context: additional context with information such as: gender, native language, hierarchy relationship and type of relationship type. 
\end{enumerate}

For each of these context treatments, each email went through two iteration related treatment --- where, half of the time, email went through two iterations of improvement; while other half of the time, email went through three iterations of improvement. The third iteration was executed during the consensus phase and caused negligible effect. Overall, for each treatment, the email went through two phases --- tone scaffolding and the consensus process --- and received revision input from six crowd workers.

To filter out non-English speakers from our crowd workers, we chose only individuals located in the US. We also only chose individuals with a 95\% approval rating. Upon inquiring about the selected workers, we found that 58\% were male, 99\% were native English speakers, and 100\% assessed themselves to be expert and fluent in English. Of these workers, 31\% had a high school degree, 51\% had an undergraduate college degree, and 17\% had a graduate degree.

\subsection{Step 4: Evaluate CrowdTone output through people who shared emails (recipients)
}

To assess the ability of CrowdTone to improve email tone, we reached out to our participating recipients again, and asked them to assess the CrowdTone-generated emails in terms of the tone appropriateness and improvement in quality of tone. In this second round of interaction, we received responses for 24 of the original 29 emails. Five participants in the initial group were not available to provide feedback on the newer versions of the emails they submitted. 

\subsection{Step 5: Evaluate CrowdTone output through professionals for further validation}

To make up for the participants lost in the second round of interaction with the recipients, and to get a second perspective, we recruited three professional writers from Upwork. These individuals had ratings of 4.5 or higher, were from North America, had completed more than 100 hours of work on the Upwork platform, and assessed themselves to be experts in English.

The professionals were asked to fill in a similar survey as participating recipients, where they were shown an original email with its context, followed by newer versions. Professionals were asked to rate the tone appropriateness and whether the newer versions were improved, and by how much. We derived consensus from their responses that helped us make decisions about the quality of work produced by CrowdTone.

\section{RESULTS AND DISCUSSION}
 
In this section, we discuss the results of our CrowdTone evaluation. As the goal for this study was to evaluate the quality of CrowdTone output --- for each sample email, we received feedback from recipients and professionals via survey. These individuals compared this version to the original. After reviewing this feedback, we came to the following conclusions. 

\subsection{CrowdTone generates emails with appropriate tone}

\begin{figure}[t]
\centering
  \includegraphics[width=0.9\columnwidth]{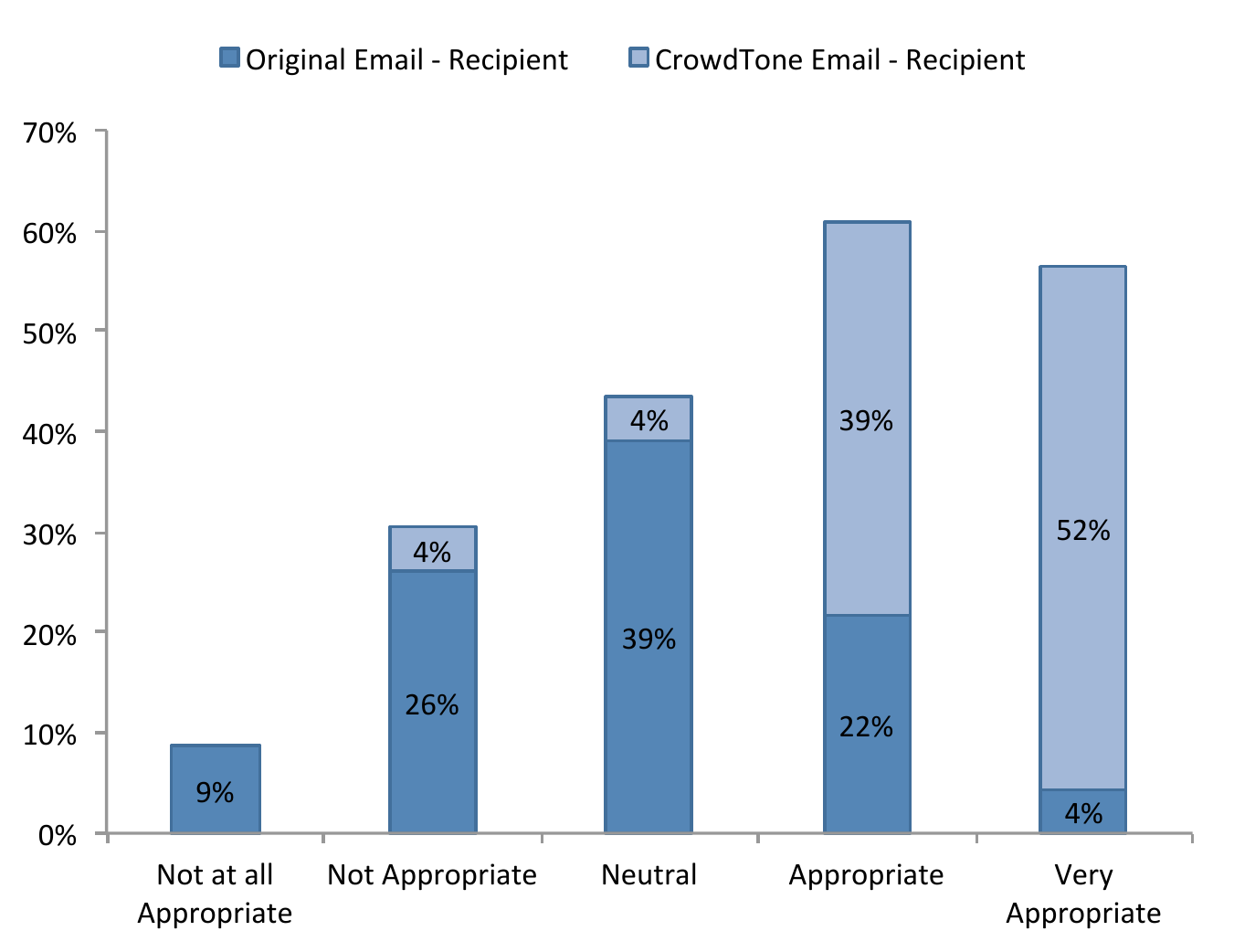}
  \caption{Appropriateness of tone for emails generated from CrowdTone compared to original emails --- as rated by the recipients}~\label{fig:approreci}
\end{figure}

According to the recipients' feedback, the original emails were mostly inappropriate. Only 26\% were judged to be ``appropriate'' or ``very appropriate''. As Figure~\ref{fig:approreci} shows, after the CrowdTone process, the percentage of ``appropriate'' and ``very appropriate'' emails rose to 91\%, with the majority considered to be ``very appropriate''. No emails from CrowdTone were assessed to be ``not at all appropriate.''

\begin{figure}[t]
\centering
  \includegraphics[width=0.9\columnwidth]{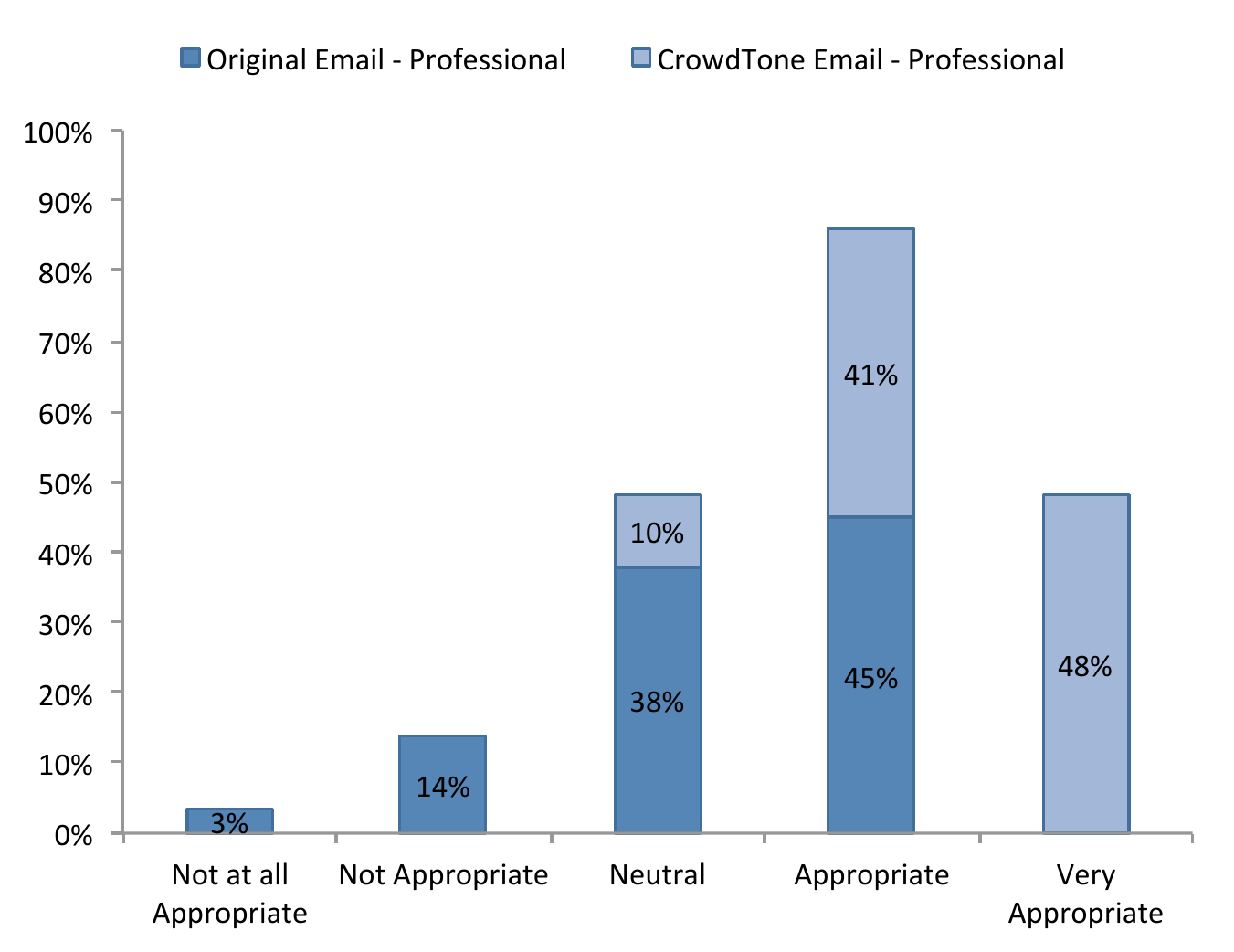}
  \caption{Appropriateness of tone for emails generated from CrowdTone compared to original emails --- as rated by the professionals}~\label{fig:approprof}
\end{figure}

According to the professionals' feedback the percentage of ``appropriate'' and ``very appropriate'' emails rose from 45\% for the originals to 90\% for the CrowdTone emails. No emails from CrowdTone were judged by the professionals to be ``inappropriate'' (see Figure~\ref{fig:approprof}).

Overall, we found CrowdTone produced emails that were both appropriate in tone and improvements over the originals.

\subsection{CrowdTone produces high-quality emails with or without additional context}

Besides extra work to provide additional context, a lot of times users may not have more contextual information of the recipient --- such as: hierarchy relationship, gender and native language. In this light, we wanted our system to be able to produce high-quality results, with or without additional context --- and hence to test whether additional context substantially improves the quality. That is, whether the CrowdTone results are substantially affected by the amount of context provided.

\begin{figure}[t]
\centering
  \includegraphics[width=0.9\columnwidth]{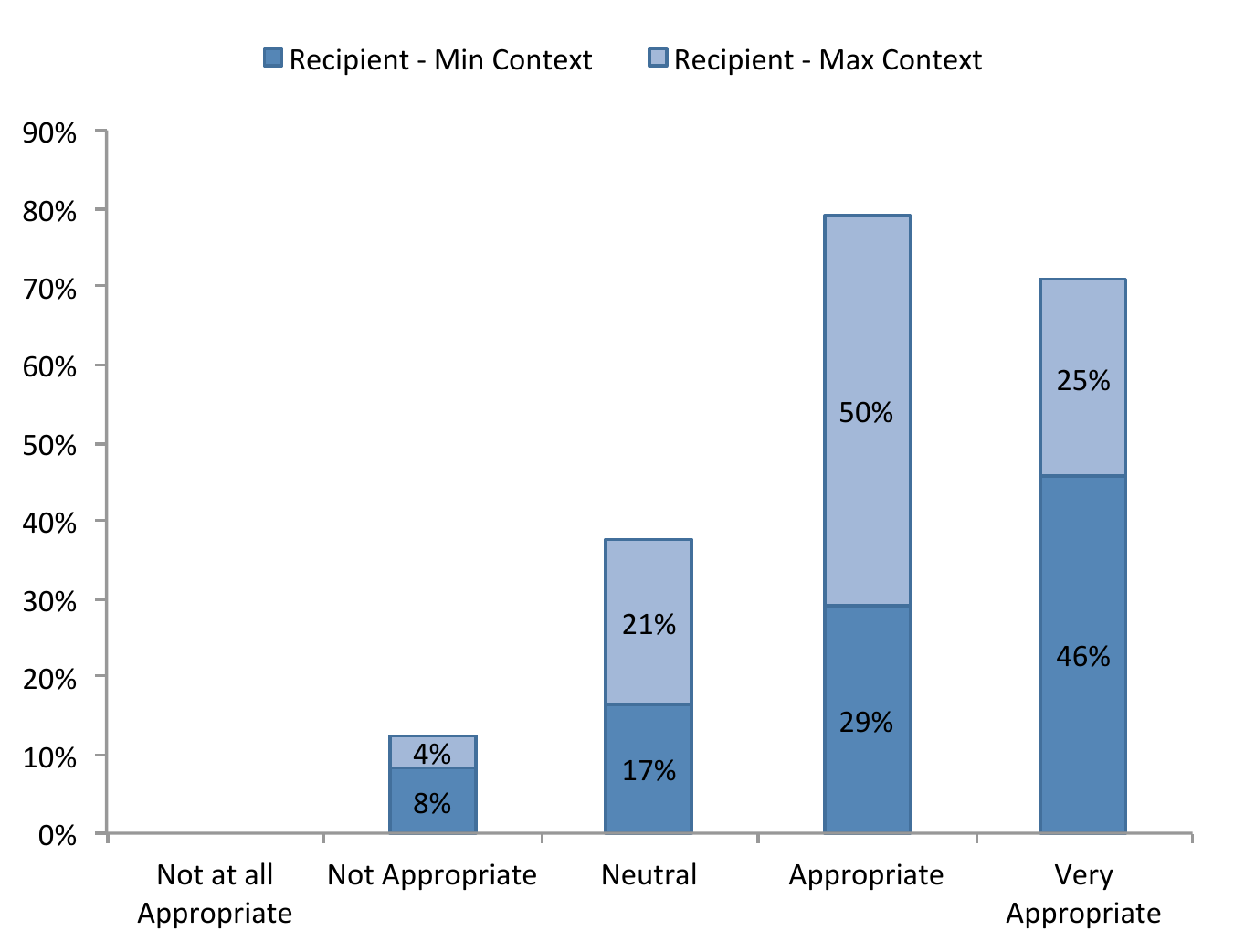}
  \caption{Appropriateness of tone for emails generated from CrowdTone using minimal and maximum context --- as rated by the recipients}~\label{fig:approreciab}
\end{figure}

As Figure~\ref{fig:approreciab} shows, the evaluation recipients designated 29\% of emails with minimal context (mandatory) input to be ``appropriate'' and 46\% of these emails to be ``very appropriate'', totaling 75\%. At the same time, the recipients judged 50\% of the CrowdTone emails with maximum context (including additional information like hierarchy and relationship) to be ``appropriate'' and 25\% to be ``very appropriate'' emails, also totaling 75\%. From this, we learned that the total percentage of CrowdTone emails judged to be ``appropriate'' and ``very appropriate'' is the same for minimal and maximum inputs.

\begin{figure}[t]
\centering
  \includegraphics[width=0.9\columnwidth]{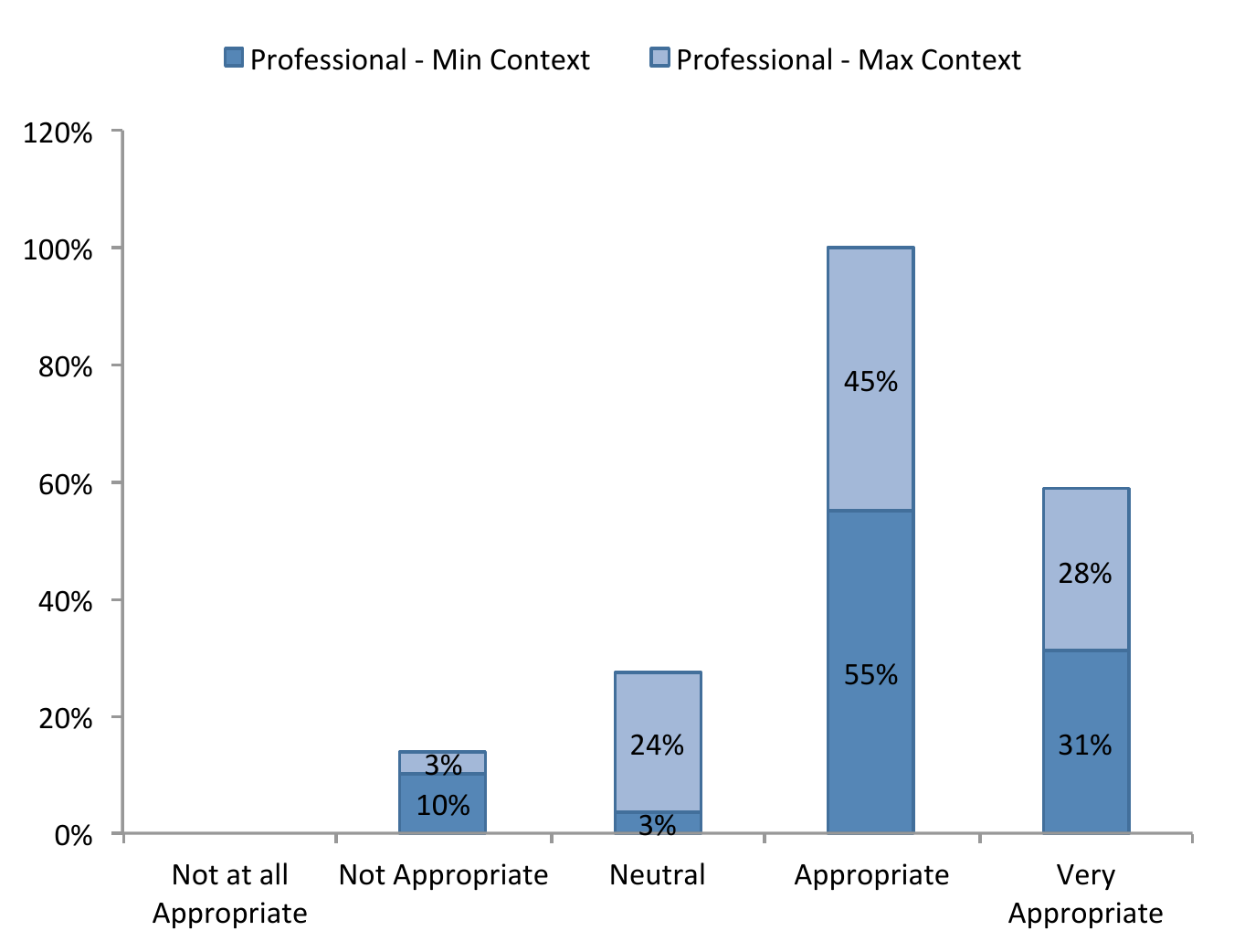}
  \caption{Appropriateness of tone for emails generated from CrowdTone using minimal and maximum context --- as rated by the professionals}~\label{fig:approprofab}
\end{figure}

As Figure~\ref{fig:approprofab} illustrates, the professional writers found 55\% of CrowdTone emails with minimal context to be ``appropriate'' and 31\% to be ``very appropriate'', totaling 86\%. In contrast, they found 45\% of emails with maximum context input to be ``appropriate'' and 28\% to be ``very appropriate'', totaling 72\%.

After conducting a blind ranking of emails produced using minimal and maximum context input, we found that participating recipients and professionals ranked output from maximum context emails higher --- 54\% and 61\% of the time, respectively. This is not substantially different from the emails with minimal context inputs.

Overall, we learned that adding additional context does not create a substantial difference in output --- CrowdTone's tone scaffolding and consensus workflow is robust with regard to contextual information.

\subsection{CrowdTone substantially improves the quality of tone over the original emails}

As part of the same survey, participating recipients were asked to quantify the improvement in the newer email's tone on a Likert scale from ``not at all improved'' to ``significant improvement''.

\begin{figure}[t] 
\centering
  \includegraphics[width=0.9\columnwidth]{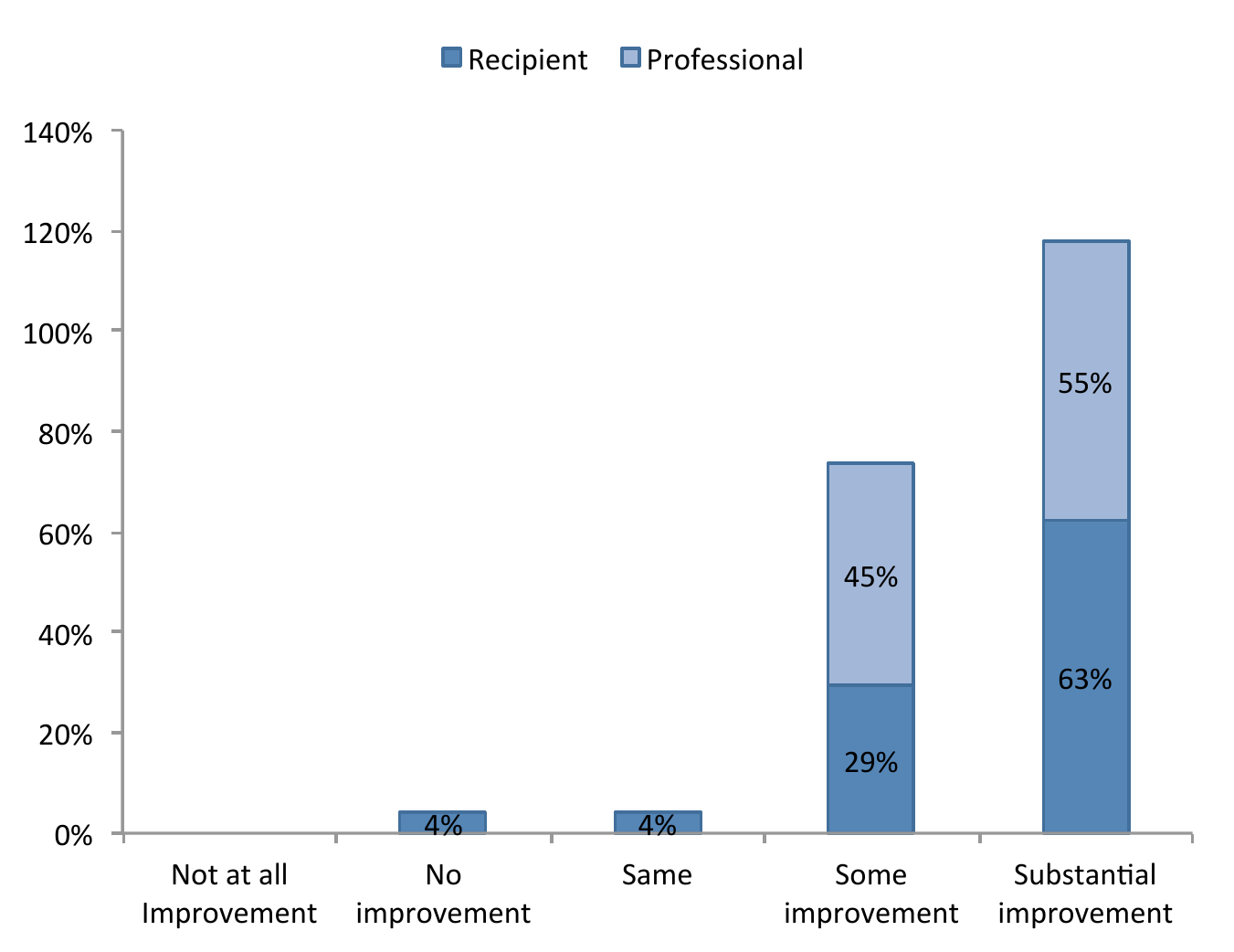}
  \caption{CrowdTone improvement over the original email --- as rated by the recipients and professionals}~\label{fig:signi}
\end{figure}

As Figure~\ref{fig:signi} shows, the recipients observed significant improvement in 63\% and some improvement in 29\% of the emails, totaling 92\%. Recipients observed no improvement for 4\% of emails that had already been deemed ``appropriate''.

The averaged responses of the three professionals indicated that 55\% of the emails showed significant improvement and that 45\% showed some improvement, totaling to 100\%.

\subsection{CrowdTone generates a professional-quality tone that matches the writing expectations of recipient}

To understand whether CrowdTone produces high-quality emails, that is, those that meet the writing expectations of recipients or are ``professional'' in quality, we asked the survey respondents whether the newer emails were of expert quality or would match the quality of an email they had written. They responded on a Likert scale that ranged from ``totally disagree'' to ``totally agree''. 

\begin{figure}[t]
\centering
  \includegraphics[width=0.9\columnwidth]{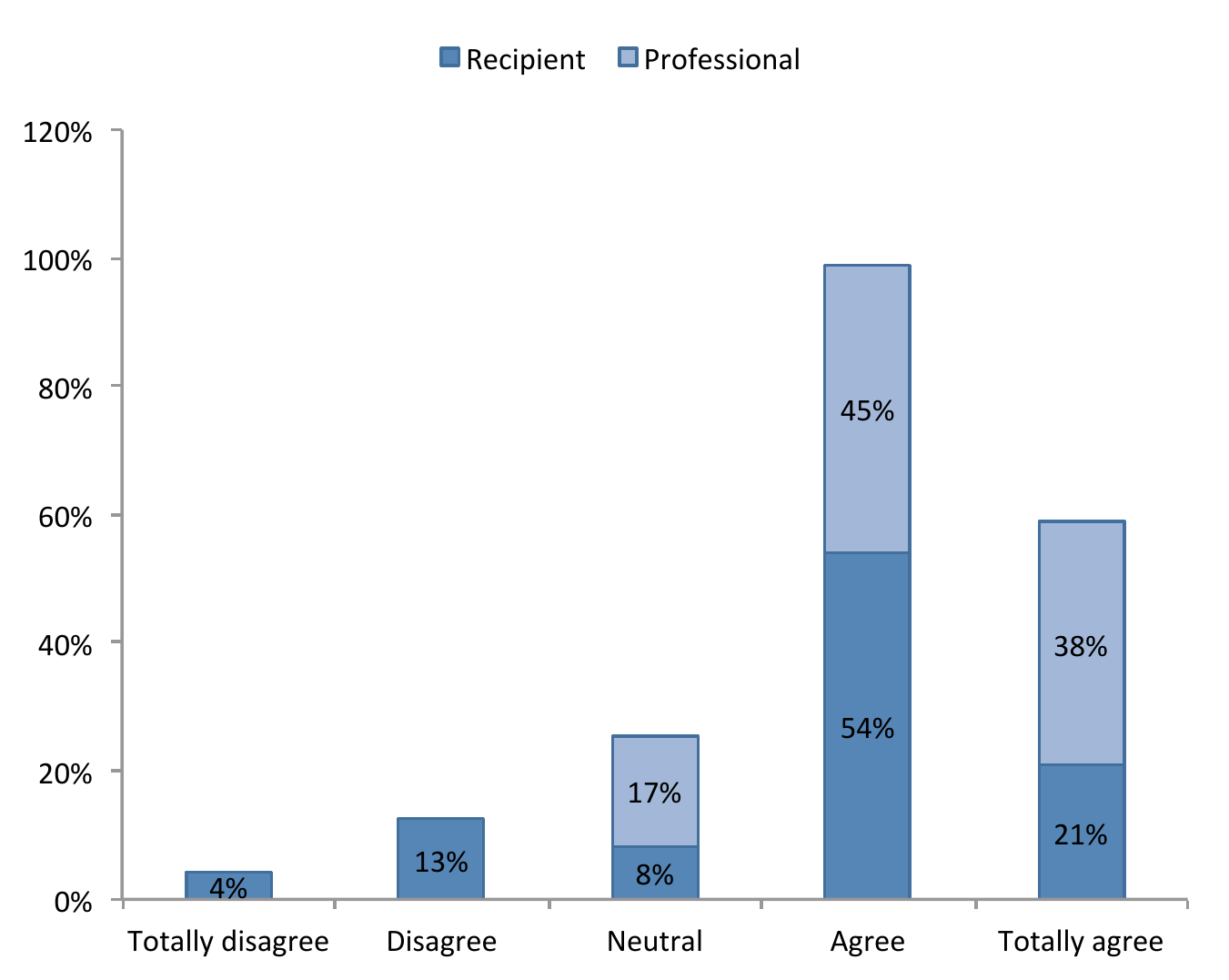}
  \caption{Agreement to  high quality  matching that of professionals --- as rated by the recipients and professionals}~\label{fig:agree}
\end{figure}

As Figure~\ref{fig:agree} illustrates, the recipients responded that they ``totally agreed'' regarding 21\% of the emails and ``agreed'' regarding the high quality of 54\% of the CrowdTone emails, totaling 75\%. The professionals indicated that they ``totally agreed'' regarding high quality for 38\% of the emails and ``agreed'' for 45\%, totaling 83\%. Together, the professionals and recipients found that the quality of the newer emails matched that of professional work.

\subsection{CrowdTone increases the chances of hearing back from the recipients}

One aspect that often determines the ``success'' of an email is whether it elicits a response. Although recipients can have multiple reasons for replying to an email, we focused here on tone as the primary motivator. Specifically, we investigated whether recipients responded to the original emails and whether they would respond to the newer CrowdTone versions. 

\begin{figure}[t]
\centering
  \includegraphics[width=0.9\columnwidth]{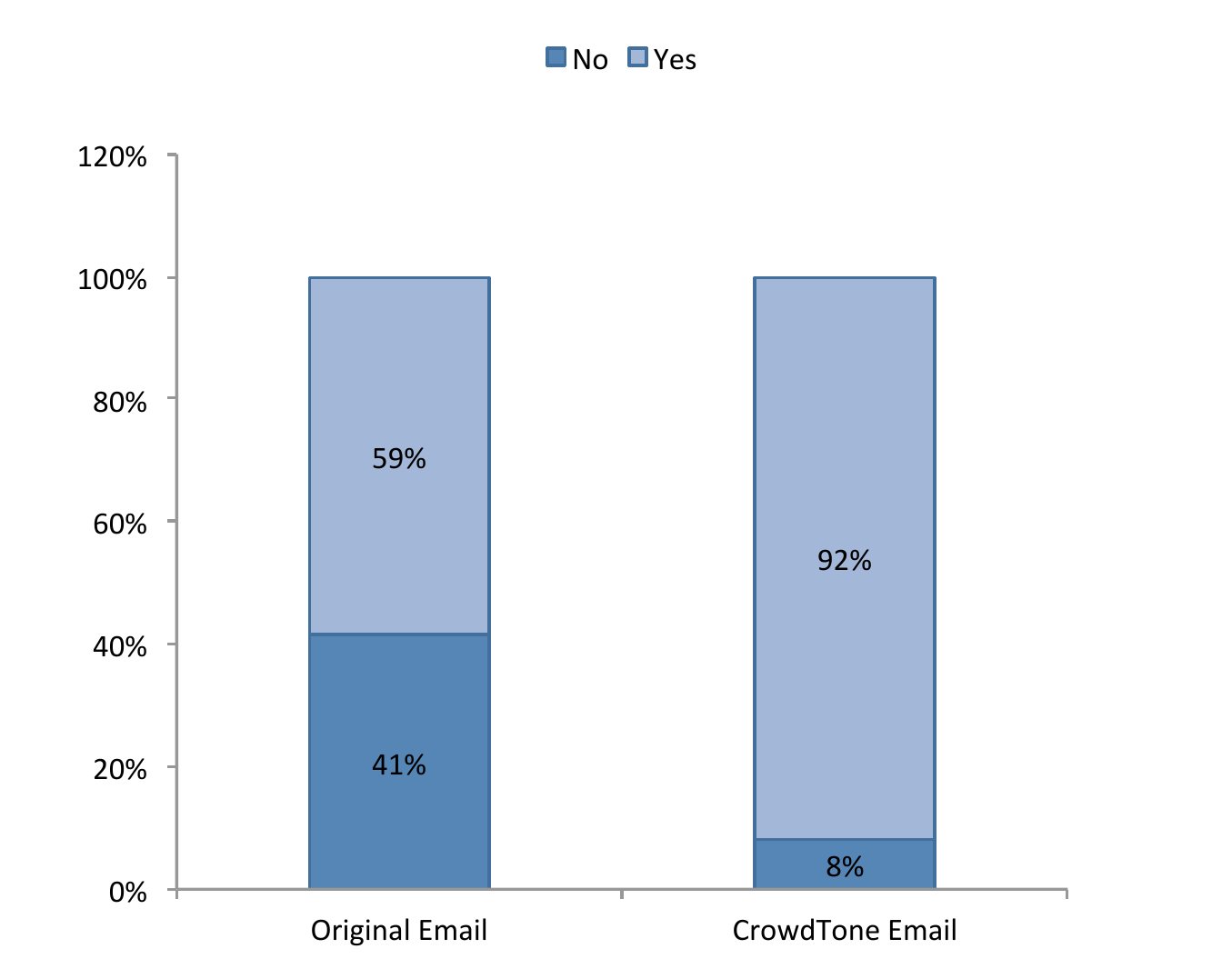}
  \caption{Chances of responding back --- as rated by the recipients}~\label{fig:respondback}
\end{figure}

As Figure~\ref{fig:respondback} shows, we found that recipients did not respond to 41\% of the original emails. With the emails generated by CrowdTone, the recipients judged that they would not respond to just 8\% of the emails. In other words they would respond to 92\% of the CrowdTone emails.

\subsection{The crowd behind the CrowdTone found the tone-scaffolding process easy and effective in accomplishing their task}

The crowd is essential to CrowdTone. To help us get better results and to optimize the crowd workers' experience, we incorporated tone scaffolding and consensus workflow into CrowdTone. After recruiting crowd workers that satisfied minimal requirements --- a 95\% approval rating and geographical location in the United States --- we got high-quality results that pleased the email recipients and professional writers. To help us understand the crowd's experience, we asked the workers to give us feedback as part of the HIT. 

Through a preliminary qualitative study, we found that the crowd deemed the CrowdTone process to be extremely effective. It helped them identify and improve email tone through multiple iterations. Most crowd workers also acknowledged their enjoyment of the HITs and asked to do more, as the quotes below demonstrate.  

\textit{``The HIT was easy enough. I understood what was being asked of me, and I was able to do the task quickly.''}

\textit{``HIT was very easy. I don't see any way you could help me make a tone related decision faster with this specific HIT.''}

\textit{``The HIT was far more easy and straightforward than I expected. Please post more.
''}

\section{CONCLUSIONS AND FUTURE WORK}

This paper presents CrowdTone, a crowd-powered system that receives email content and context, and outputs the same email with improved tone. CrowdTone utilizes crowd workers from Amazon Mechanical Turk, who go through specifically designed tone scaffolding and consensus phases --- from self-identifying tone related requirements to fixing it --- without user instructions. To evaluate the quality of the enhanced emails produced by CrowdTone, we collected 29 emails from 22 participants, and had them reviewed by said participating recipients and three professional writers from Upwork.  Overall, CrowdTone's core contributions are:
\begin{itemize}
  \item Robustness with regard to the presence or absence of context provided. 
  \item Substantial improvements in the quality of tone over the original email. Based on the participating recipients' evaluation, 60\% or more of the emails processed exhibited substantial improvement.
  \item Generation of professional-quality tone that in the evaluation matched the writing expectations of the recipients. When surveyed, 75\% of these participants ``agreed'' or ``totally agreed'' when asked whether the CrowdTone emails were of high quality.
  \item Improved chances of getting a response back from email recipients. The participating recipients indicated that they would respond to more than 90\% of the emails, a substantial improvement from the original 59\%.
  \item Tone-scaffolding process was reported as  easy and effective by crowdworkers. 
\end{itemize}

Through CrowdTone, we extend existing crowd-powered systems research for efficient writing solutions, and present a novel process that self-identifies and improves email tone. That said, the system also showed several limitations that are seeds for future work:

\begin{itemize}
\item Privacy. Participants raised concerns about the privacy implications of sharing private communication with crowdworkers without all the parties being aware of it. Future work might explore ways of addressing this through obfuscation and explicit norm setting, and other mechanisms explored in similar prior work~\cite{Kokkalis:2013:EME:2441776.2441922}
\item Delay. Having to wait for the process of tone setting to go through introduces delays that might not be suitable for some users, and some scenarios. Optimizing for real-time responses through crowd-retainers as it was done in similar research~\cite{Bernstein:2011:CTS:2047196.2047201} is something worth exploring.
\item Deep integration. At the moment, CrowdTone is not deeply embedded in the regular workflow of sending emails. Creating a plugin could be one way of addressing this, however, plugins are not universally used, for example, mobile email clients rarely allow for plugins.  However, deeper integration can provide access to conversation history --- making CrowdTone more intelligent in its ability to determine deeper context and meaning. More work is needed to understand the best mechanisms for integration into people's work styles. 

\item Restricted applications. Though CrowdTone can be utilized to improve the tone of any type of written communication, the current paper focuses on its application for professional emails. In future, we plan to apply the core of our existing approach and expand it to different use cases beyond emails: to other types of written communications, domains or media. 
\end{itemize}

We see a lot of potential for introducing third-party feedback and fixing into our professional communication. Large companies often have marketing groups in charge of corporate communication, but we envision this level of attention to detail could be affordable for individuals. More generally we see this work as yet another example of a potential for hybrid intelligence where crowds and AI can come together to optimize processes and leverage the best of both worlds.

\Urlmuskip=0mu plus 1mu\relax
\bibliographystyle{SIGCHI-Reference-Format}
\bibliography{proceedings}


\begin{thebibliography}{00}


\ifx \showCODEN    \undefined \def \showCODEN     #1{\unskip}     \fi
\ifx \showDOI      \undefined \def \showDOI       #1{{\tt DOI:}\penalty0{#1}\ }
  \fi
\ifx \showISBNx    \undefined \def \showISBNx     #1{\unskip}     \fi
\ifx \showISBNxiii \undefined \def \showISBNxiii  #1{\unskip}     \fi
\ifx \showISSN     \undefined \def \showISSN      #1{\unskip}     \fi
\ifx \showLCCN     \undefined \def \showLCCN      #1{\unskip}     \fi
\ifx \shownote     \undefined \def \shownote      #1{#1}          \fi
\ifx \showarticletitle \undefined \def \showarticletitle #1{#1}   \fi
\ifx \showURL      \undefined \def \showURL       #1{#1}          \fi

\bibitem{alchemy}
{IBM Alchemy}. 2015.
\newblock Alchemy Sentiment Analysis API.
\newblock   (2015).
\newblock
\newblock
\shownote{\url{http://www.alchemyapi.com/api/sentiment-analysis}.}


\bibitem{microsoft}
{Microsoft Azure}. 2015.
\newblock Lexicon Based Sentiment Analysis.
\newblock   (2015).
\newblock
\newblock
\shownote{\url{https://azure.microsoft.com}.}


\bibitem{Bernstein:2011:CTS:2047196.2047201}
{Michael~S. Bernstein}, {Joel Brandt}, {Robert~C. Miller}, {and} {David~R.
  Karger}. 2011.
\newblock \showarticletitle{Crowds in Two Seconds: Enabling Realtime
  Crowd-powered Interfaces}. In {\em Proceedings of the 24th Annual ACM
  Symposium on User Interface Software and Technology} {\em (UIST '11)}. ACM,
  New York, NY, USA, 33--42.
\newblock
\showISBNx{978-1-4503-0716-1}
\showDOI{%
\url{http://dx.doi.org/10.1145/2047196.2047201}}


\bibitem{Bernstein:2010:SWP:1866029.1866078}
{Michael~S. Bernstein}, {Greg Little}, {Robert~C. Miller}, {Bj\"{o}rn
  Hartmann}, {Mark~S. Ackerman}, {David~R. Karger}, {David Crowell}, {and}
  {Katrina Panovich}. 2010.
\newblock \showarticletitle{Soylent: A Word Processor with a Crowd Inside}. In
  {\em Proceedings of the 23Nd Annual ACM Symposium on User Interface Software
  and Technology} {\em (UIST '10)}. ACM, New York, NY, USA, 313--322.
\newblock
\showISBNx{978-1-4503-0271-5}
\showDOI{%
\url{http://dx.doi.org/10.1145/1866029.1866078}}


\bibitem{byron2008carrying}
{Kristin Byron}. 2008.
\newblock \showarticletitle{Carrying too heavy a load? The communication and
  miscommunication of emotion by email}.
\newblock {\em Academy of Management Review\/} {33}, 2 (2008), 309.
\newblock


\bibitem{cai2016chain}
{Carrie~J Cai}, {Shamsi~T Iqbal}, {and} {Jaime Teevan}. 2016.
\newblock \showarticletitle{Chain Reactions: The Impact of Order on Microtask
  Chains}. In {\em Proceedings of CHI}.
\newblock


\bibitem{Dow:2011:PDS:1978942.1979359}
{Steven Dow}, {Julie Fortuna}, {Dan Schwartz}, {Beth Altringer}, {Daniel
  Schwartz}, {and} {Scott Klemmer}. 2011.
\newblock \showarticletitle{Prototyping Dynamics: Sharing Multiple Designs
  Improves Exploration, Group Rapport, and Results}. In {\em Proceedings of the
  SIGCHI Conference on Human Factors in Computing Systems} {\em (CHI '11)}.
  ACM, New York, NY, USA, 2807--2816.
\newblock
\showISBNx{978-1-4503-0228-9}
\showDOI{%
\url{http://dx.doi.org/10.1145/1978942.1979359}}


\bibitem{Dow:2012:SCY:2145204.2145355}
{Steven Dow}, {Anand Kulkarni}, {Scott Klemmer}, {and} {Bj\"{o}rn Hartmann}.
  2012.
\newblock \showarticletitle{Shepherding the Crowd Yields Better Work}. In {\em
  Proceedings of the ACM 2012 Conference on Computer Supported Cooperative
  Work} {\em (CSCW '12)}. ACM, New York, NY, USA, 1013--1022.
\newblock
\showISBNx{978-1-4503-1086-4}
\showDOI{%
\url{http://dx.doi.org/10.1145/2145204.2145355}}


\bibitem{Dow:2010:PPL:1879831.1879836}
{Steven~P. Dow}, {Alana Glassco}, {Jonathan Kass}, {Melissa Schwarz},
  {Daniel~L. Schwartz}, {and} {Scott~R. Klemmer}. 2010.
\newblock \showarticletitle{Parallel Prototyping Leads to Better Design
  Results, More Divergence, and Increased Self-efficacy}.
\newblock {\em ACM Trans. Comput.-Hum. Interact.\/} {17}, 4, Article 18 (Dec.
  2010), 24 pages.
\newblock
\showISSN{1073-0516}
\showDOI{%
\url{http://dx.doi.org/10.1145/1879831.1879836}}


\bibitem{Dow:2009:EPU:1640233.1640260}
{Steven~P. Dow}, {Kate Heddleston}, {and} {Scott~R. Klemmer}. 2009.
\newblock \showarticletitle{The Efficacy of Prototyping Under Time
  Constraints}. In {\em Proceedings of the Seventh ACM Conference on Creativity
  and Cognition} {\em (C\&\#38;C '09)}. ACM, New York, NY, USA, 165--174.
\newblock
\showISBNx{978-1-60558-865-0}
\showDOI{%
\url{http://dx.doi.org/10.1145/1640233.1640260}}


\bibitem{epley2005you}
{Nicholas Epley} {and} {Justin Kruger}. 2005.
\newblock \showarticletitle{When what you type isn’t what they read: The
  perseverance of stereotypes and expectancies over e-mail}.
\newblock {\em Journal of Experimental Social Psychology\/} {41}, 4 (2005),
  414--422.
\newblock


\bibitem{google}
{Google}. 2015.
\newblock Sentiment Analysis Model.
\newblock   (2015).
\newblock
\newblock
\shownote{\url{https://cloud.google.com/prediction/docs/sentiment_analysis}.}


\bibitem{greer2016introduction}
{Nick Greer}, {Jaime Teevan}, {and} {Shamsi~T Iqbal}. 2016.
\newblock {\em An introduction to technological support for writing}.
\newblock {T}echnical {R}eport. Technical Report. Microsoft Research Tech
  Report MSR-TR-2016-001.
\newblock


\bibitem{crystal}
{Crystal~Project Inc}. 2015.
\newblock Crystal.
\newblock   (2015).
\newblock
\newblock
\shownote{\url{https://www.crystalknows.com/}.}


\bibitem{fastcompany}
{Eric Jaffe}. 2014.
\newblock Why It's So Hard To Detect Emotion In Emails And Texts.
\newblock   (2014).
\newblock
\newblock
\shownote{\url{http://www.fastcodesign.com/3036748/evidence/why-its-so-hard-to-detect-emotion-in-emails}.}


\bibitem{Kim:2014:EEC:2531602.2531638}
{Joy Kim}, {Justin Cheng}, {and} {Michael~S. Bernstein}. 2014.
\newblock \showarticletitle{Ensemble: Exploring Complementary Strengths of
  Leaders and Crowds in Creative Collaboration}. In {\em Proceedings of the
  17th ACM Conference on Computer Supported Cooperative Work \&\#38; Social
  Computing} {\em (CSCW '14)}. ACM, New York, NY, USA, 745--755.
\newblock
\showISBNx{978-1-4503-2540-0}
\showDOI{%
\url{http://dx.doi.org/10.1145/2531602.2531638}}


\bibitem{storia}
{Joy Kim} {and} {Andres Monroy-Hernandez}. 2016.
\newblock \showarticletitle{Storia: Summarizing Social Media Content Based on
  Narrative Theory Using Crowdsourcing}. In {\em Proceedings of the 19th ACM
  Conference on Computer-Supported Cooperative Work \& Social Computing} {\em
  (CSCW '16)}. ACM, New York, NY, USA, 1018--1027.
\newblock
\showISBNx{978-1-4503-3592-8}
\showDOI{%
\url{http://dx.doi.org/10.1145/2818048.2820072}}


\bibitem{Kittur:2011:CCC:2047196.2047202}
{Aniket Kittur}, {Boris Smus}, {Susheel Khamkar}, {and} {Robert~E. Kraut}.
  2011.
\newblock \showarticletitle{CrowdForge: Crowdsourcing Complex Work}. In {\em
  Proceedings of the 24th Annual ACM Symposium on User Interface Software and
  Technology} {\em (UIST '11)}. ACM, New York, NY, USA, 43--52.
\newblock
\showISBNx{978-1-4503-0716-1}
\showDOI{%
\url{http://dx.doi.org/10.1145/2047196.2047202}}


\bibitem{Kokkalis:2013:EME:2441776.2441922}
{Nicolas Kokkalis}, {Thomas K\"{o}hn}, {Carl Pfeiffer}, {Dima Chornyi},
  {Michael~S. Bernstein}, {and} {Scott~R. Klemmer}. 2013.
\newblock \showarticletitle{EmailValet: Managing Email Overload Through
  Private, Accountable Crowdsourcing}. In {\em Proceedings of the 2013
  Conference on Computer Supported Cooperative Work} {\em (CSCW '13)}. ACM, New
  York, NY, USA, 1291--1300.
\newblock
\showISBNx{978-1-4503-1331-5}
\showDOI{%
\url{http://dx.doi.org/10.1145/2441776.2441922}}


\bibitem{kruger2005egocentrism}
{Justin Kruger}, {Nicholas Epley}, {Jason Parker}, {and} {Zhi-Wen Ng}. 2005.
\newblock \showarticletitle{Egocentrism over e-mail: can we communicate as well
  as we think?}
\newblock {\em Journal of personality and social psychology\/} {89}, 6 (2005),
  925.
\newblock


\bibitem{Kulkarni:2012:CCW:2145204.2145354}
{Anand Kulkarni}, {Matthew Can}, {and} {Bj\"{o}rn Hartmann}. 2012.
\newblock \showarticletitle{Collaboratively Crowdsourcing Workflows with
  Turkomatic}. In {\em Proceedings of the ACM 2012 Conference on Computer
  Supported Cooperative Work} {\em (CSCW '12)}. ACM, New York, NY, USA,
  1003--1012.
\newblock
\showISBNx{978-1-4503-1086-4}
\showDOI{%
\url{http://dx.doi.org/10.1145/2145204.2145354}}


\bibitem{Lasecki:2011:RCC:2047196.2047200}
{Walter~S. Lasecki}, {Kyle~I. Murray}, {Samuel White}, {Robert~C. Miller},
  {and} {Jeffrey~P. Bigham}. 2011.
\newblock \showarticletitle{Real-time Crowd Control of Existing Interfaces}. In
  {\em Proceedings of the 24th Annual ACM Symposium on User Interface Software
  and Technology} {\em (UIST '11)}. ACM, New York, NY, USA, 23--32.
\newblock
\showISBNx{978-1-4503-0716-1}
\showDOI{%
\url{http://dx.doi.org/10.1145/2047196.2047200}}


\bibitem{Lasecki:2013:CCC:2501988.2502057}
{Walter~S. Lasecki}, {Rachel Wesley}, {Jeffrey Nichols}, {Anand Kulkarni},
  {James~F. Allen}, {and} {Jeffrey~P. Bigham}. 2013.
\newblock \showarticletitle{Chorus: A Crowd-powered Conversational Assistant}.
  In {\em Proceedings of the 26th Annual ACM Symposium on User Interface
  Software and Technology} {\em (UIST '13)}. ACM, New York, NY, USA, 151--162.
\newblock
\showISBNx{978-1-4503-2268-3}
\showDOI{%
\url{http://dx.doi.org/10.1145/2501988.2502057}}


\bibitem{Luther:2015:SAE:2675133.2675283}
{Kurt Luther}, {Jari-Lee Tolentino}, {Wei Wu}, {Amy Pavel}, {Brian~P. Bailey},
  {Maneesh Agrawala}, {Bj\"{o}rn Hartmann}, {and} {Steven~P. Dow}. 2015.
\newblock \showarticletitle{Structuring, Aggregating, and Evaluating
  Crowdsourced Design Critique}. In {\em Proceedings of the 18th ACM Conference
  on Computer Supported Cooperative Work \&\#38; Social Computing} {\em (CSCW
  '15)}. ACM, New York, NY, USA, 473--485.
\newblock
\showISBNx{978-1-4503-2922-4}
\showDOI{%
\url{http://dx.doi.org/10.1145/2675133.2675283}}


\bibitem{tonechecker}
{Lymbix}. 2015.
\newblock ToneCheck.
\newblock   (2015).
\newblock
\newblock
\shownote{\url{http://tonecheck.com/}.}


\bibitem{voiceandtone}
{MailChimp}. 2015.
\newblock Voice\&Tone.
\newblock   (2015).
\newblock
\newblock
\shownote{\url{http://voiceandtone.com/}.}


\bibitem{nebeling2016wearwrite}
{Michael Nebeling}, {Alexandra To}, {Anhong Guo}, {Adrian~A de Freitas}, {Jaime
  Teevan}, {Steven~P Dow}, {and} {Jeffrey~P Bigham}. 2016.
\newblock \showarticletitle{WearWrite: Crowd-Assisted Writing from
  Smartwatches}. In {\em Proceedings of CHI}.
\newblock


\bibitem{Pearl}
{Lisa Pearl} {and} {Mark Steyvers}. 2013.
\newblock \showarticletitle{“C’mon – You Should Read This”: Automatic
  Identification of Tone from Language Text}.
\newblock {\em International Journal of Computational Linguistics (IJCL)\/}
  {4}, 1 (2013), 12 -- 30.
\newblock
\showISSN{2180-1266}


\bibitem{emailnumber}
{Craig Smith}. 2016.
\newblock By the Numbers: 60 Incredible Email Statistics.
\newblock   (2016).
\newblock
\newblock
\shownote{\url{http://expandedramblings.com/index.php/email-statistics/}.}


\bibitem{liwc}
{Yla~R. Tausczik} {and} {James~W. Pennebaker}. 2010.
\newblock \showarticletitle{The Psychological Meaning of Words: LIWC and
  Computerized Text Analysis Methods}.
\newblock {\em Journal of Language and Social Psychology\/} {29}, 1 (2010),
  24--54.
\newblock
\showDOI{%
\url{http://dx.doi.org/10.1177/0261927X09351676}}


\bibitem{teevan2016supporting}
{Jaime Teevan}, {Shamsi Iqbal}, {and} {Curtis von Veh}. 2016.
\newblock \showarticletitle{Supporting Collaborative Writing with Microtasks}.
  In {\em Proceedings of CHI}.
\newblock


\bibitem{ibm}
{IBM Watson}. 2015.
\newblock Tone Analyzer.
\newblock   (2015).
\newblock
\newblock
\shownote{\url{http://www.ibm.com/smarterplanet/us/en/ibmwatson/developercloud/tone-analyzer.html}.}


\bibitem{ego}
{Lea Winerman}. 2006.
\newblock \showarticletitle{{E-mails and egos: An inability to step outside of
  one's own head may be behind e-mail miscommunication, according to recent
  research.}}
\newblock {\em American Psychological Association\/} {37}, 2 (2006), 16.
\newblock


\bibitem{wordzen}
{Inc Wordzen}. 2015.
\newblock Wordzen.
\newblock   (2015).
\newblock
\newblock
\shownote{\url{http://www.wordzen.com/}.}


\bibitem{egomore}
{Business~Coaching Worldwide}. 2007.
\newblock Did You Know: That email recipients correctly interpret the intended
  tone of an email message only 50\% of the time?
\newblock   (2007).
\newblock
\newblock
\shownote{\url{http://www.wabccoaches.com/bcw/2007_v3_i1/didyouknow.html}.}


\bibitem{howtotone}
{WriteExpress}. 2015.
\newblock How to use tone in your writing.
\newblock   (2015).
\newblock
\newblock
\shownote{\url{http://www.writeexpress.com/tone.html}.}


\end{thebibliography}

\end{document}